\documentclass[conference,compsoc]{IEEEtran}
\ifCLASSOPTIONcompsoc
  \usepackage[nocompress]{cite}
\else
  \usepackage{cite}
\fi
\usepackage{graphicx}
\ifCLASSINFOpdf
\else
\fi
\newcommand{\zipzap}{ZipZap}
\newcommand{\zaps}{\textit{Zaps}}
\newcommand{\zap}{\textit{Zap}}
\newcommand{\heavy}{Heavyweight}
\newcommand{\feather}{Featherweight}
\newcommand{\light}{Lightweight}
\newcommand{\noweight}{Weightless}
\usepackage{subcaption}
\usepackage{array}
\usepackage{tabularx}
\usepackage{listings}
\usepackage{multirow}
\usepackage{xcolor}

\lstdefinestyle{mystyle}{
	basicstyle=\ttfamily\footnotesize,
	breakatwhitespace=false,         
	breaklines=true,                 
	captionpos=b,                    
	keepspaces=true,                 
	numbers=left,                    
	numbersep=5pt,                  
	showspaces=false,                
	showstringspaces=false,
	showtabs=false,                  
	tabsize=2
}
\begin{document}
%
\title{\zipzap: A Blockchain Solution for Local Energy Trading}


 \author{\IEEEauthorblockN{Mario Felipe Munoz\IEEEauthorrefmark{1},
 		Kaiwen Zhang\IEEEauthorrefmark{1} and
 		Fatima Amara\IEEEauthorrefmark{2} 
 	}
 	\IEEEauthorblockA{\IEEEauthorrefmark{1}
	École de technologie supérieure, Montréal (Canada)\\
	Montreal, Quebec H3C 1K3\\
	\IEEEauthorrefmark{2}Hydro-Québec IREQ, Shawanigan (Canada)\\
 	}}

\IEEEoverridecommandlockouts

\maketitle
\begin{abstract}
\label{sec:abs}

In the last few years, electric utility companies have increasingly invested into transactive energy systems. This trend was primarily caused by the integration of distributed energy resources (DERs) and internet-of-things (IoT) devices into their existing distribution networks. Influenced by the general interest in blockchain technologies, many industry specialists are considering new, more efficient peer-to-peer market structures for DERs. Since blockchain-based energy exchanges can automate transactions between their members and provide increased levels of security thanks to smart contracts, these new initiatives may eventually revolutionize how customers interact with utility companies. In this paper, we explore the trade-off between cost and traceability in the form of on-chain and off-chain solutions. We also propose \zipzap, a first step towards a blockchain-based local smart grid system. \zipzap~is an ERC-1155 compliant solution with four different prototypes: \heavy, \feather, \light~and \noweight. The first three prototypes were developed in Solidity and deployed using Ethereum. \heavy~is fully on-chain, whereas \feather~and \light~showcase various levels of hybridization. \noweight, in turn, was deployed using Quorum, a gas-free alternative to Ethereum. Our evaluation uses realistic parameters and measures the impact of different types of metadata storage scopes, with some Ethereum prototypes showcasing gas cost reductions of more than 97\% in comparison to our fully on-chain baseline.

\end{abstract}



%
\IEEEpeerreviewmaketitle

\section{Introduction}
\label{sec:intro}

Contemporary energy markets have been greatly impacted by the falling costs of decentralized renewable energy sources (DER) and the introduction of internet of things (IoT) devices. These changes led to the creation of peer-to-peer exchanges through which renewable energy can be sold. Consequentially, new sociotechnical challenges arose in the context of energy security for both communities and individuals. One serious concern is that the many IoT devices that form smart grids generate unprecedented volumes of data---data that has to be adequately managed and protected. To makes matters worse, even though further decentralizing energy grids increases their availability, it also increases their overall attack surface area. These factors have driven industry specialists to consider blockchain systems in hopes of further automating operations and providing greater levels of data integrity.

\begin{figure*}[] 
\fbox{\parbox{\textwidth}{
\centering
	\begin{subfigure}{0.8\columnwidth}
		\centering
		\includegraphics[width=\columnwidth]{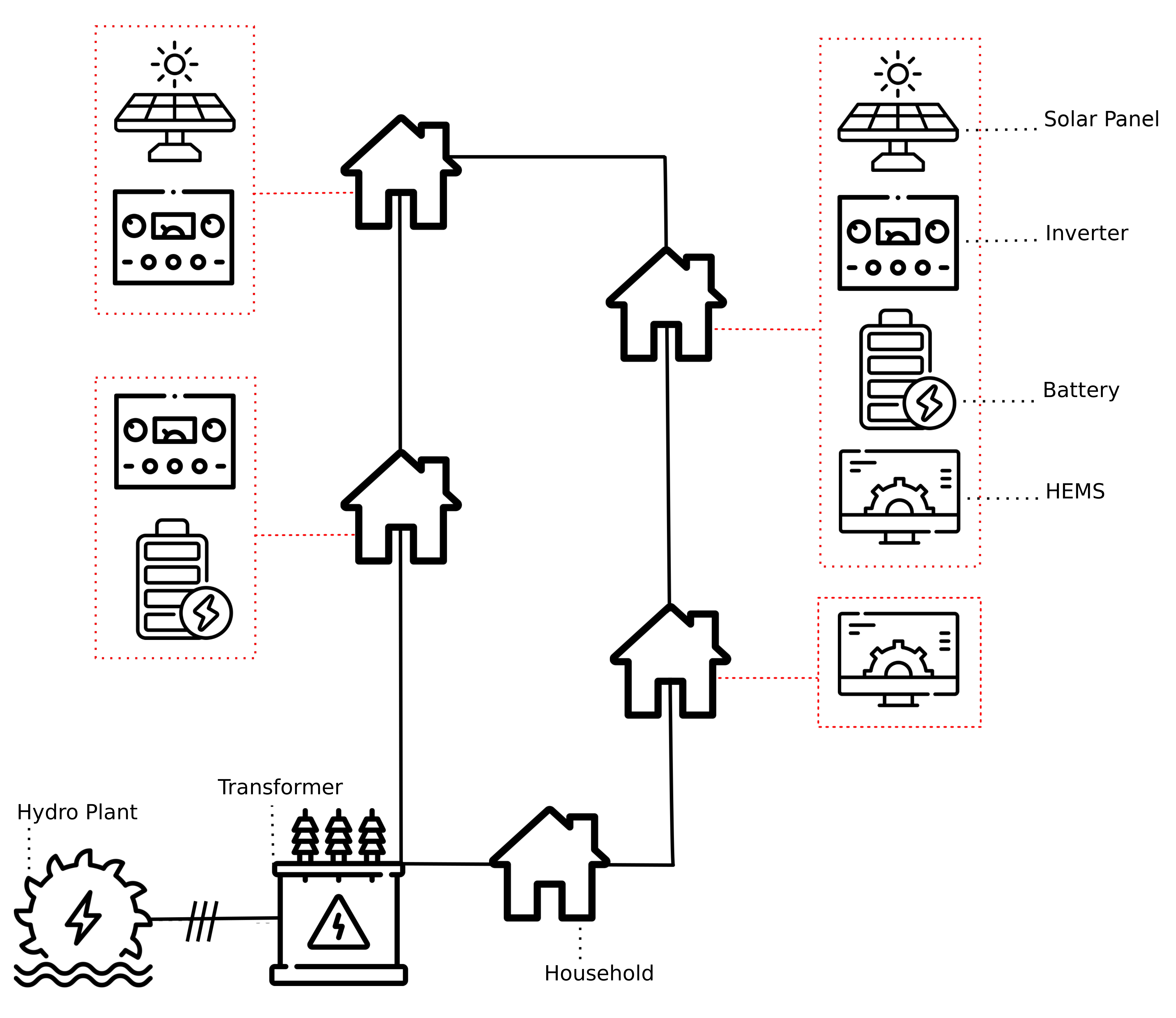}  
		\caption{}
		\label{fig:houses}
	\end{subfigure}
	\begin{subfigure}{\columnwidth}
		\centering
		\includegraphics[width=0.8\columnwidth]{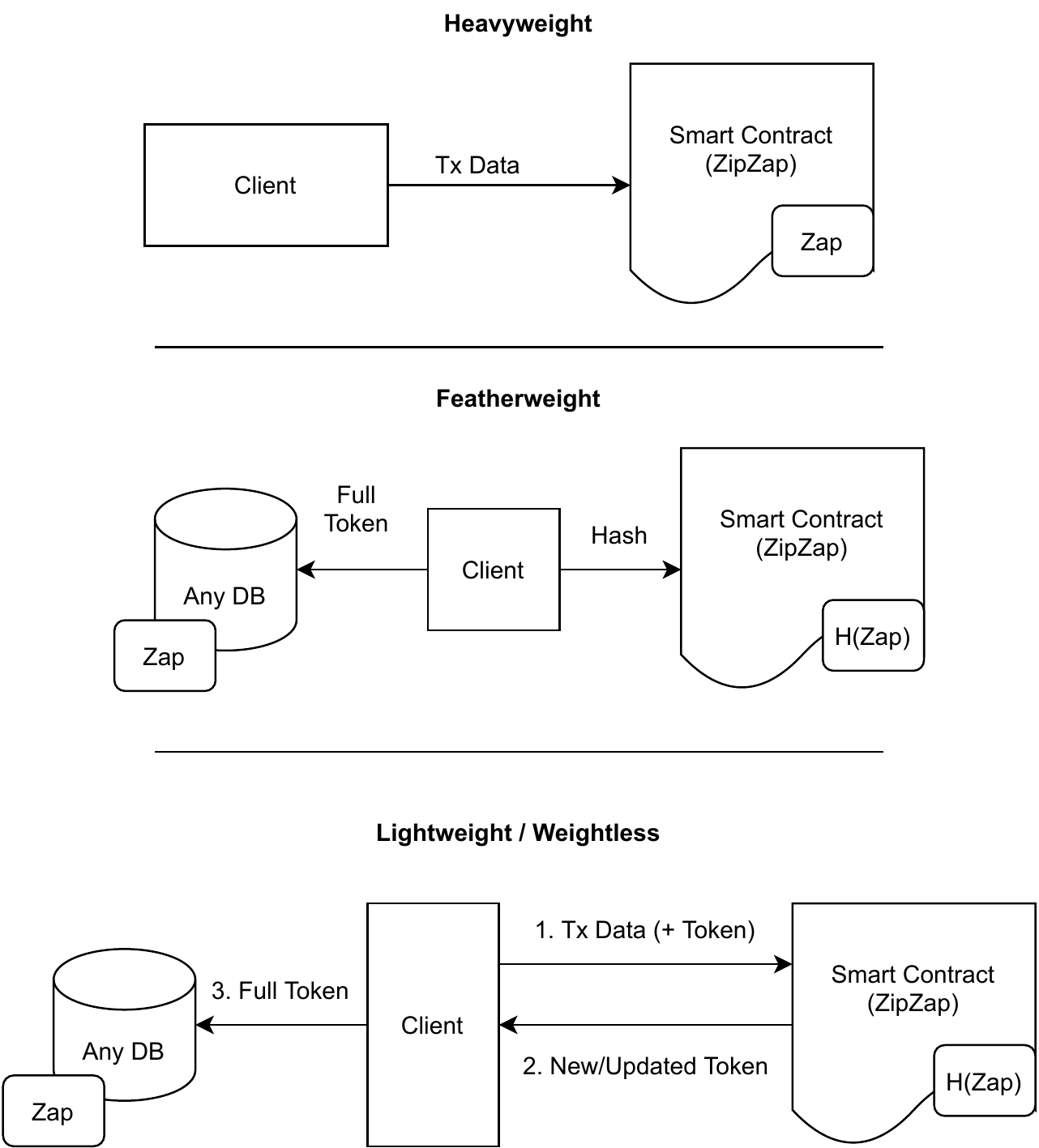}  
		\caption{}
		\label{fig:comm-diag}
	\end{subfigure}
	}}
	\caption{Neighborhood Network and Communication Diagram}
	\label{fig:together}
\end{figure*}

To meet energy industry demands, we propose \zipzap. In its current state, \zipzap~is our novel energy tokenization blockchain system for local energy exchanges, although functionality extensions are being considered. Figure \ref{fig:houses} shows our typical operational context. The utility company (represented by the hydro-power plant and transformer) supplies the basic energy infrastructure, but each household can connect to it with varying degrees of autonomy depending on their available energy equipment.

\zipzap~works by tokenizing energy quantities as they are created by different types of generators (biodiesel engines, PV cells, wind turbines, etc.). The owners of the generators can consume their own energy and sell any excess to their neighbours or to the grid. All energy-related operations are tracked by \zipzap, allowing users and auditors to verify the provenance of all energy created and spent by a household. Tokenizing energy is a noticeably harder task than tokenizing other goods such as food shipments \cite{ibm-food} because of the massive number of transactions emitted on a daily basis, and their comparatively tight latency requirements. Given the above considerations, the contributions of this study are listed as follows:

\begin{enumerate}
	\item Proposing \zipzap: an energy tokenization solution for local energy exchanges. \zipzap\ adopts a semi-fungible model for energy tokenization and is ERC-1155 compliant.
	\item Developing three Ethereum-based reference implementations of our design: \heavy, \feather~and \light. The first being a fully on-chain approach and the two latter using hybrid designs.
	\item Developing a Quorum-based reference implementation of our design.
	\item Comparing all four implementations and assessing the feasibility of our approach in terms of operating gas costs compared to energy generation costs.
	\end{enumerate}

The rest of this paper abides to a clear structure: Section \ref{sec:rw} covers previous research done in relation to blockchain systems and smart energy grids. Section \ref{sec:back} explains the value provided by such systems and some key preliminary concepts. Section \ref{sec:des} describes the composition of all four prototypes and our tokenization method. Section \ref{sec:res} shows the experimental results for all \zipzap~prototypes. Finally, Section \ref{sec:conc} ends the paper with a brief conclusion. 

\section{Related Works}
\label{sec:rw}
There is an abundance of literature published in relation to distributed ledger technology (DLT) enriched energy grids. However, we found at least one interesting discrepancy between papers. To begin with, the previous comparative works reviewed \cite{smart-examples, smart-categories} often showcase no small number of Ethereum-based solutions for several energy-related services. In a similar vein, there are also a number of other such works that describe the different requirements that each energy use case has, as well as the features and limitations of many blockchain platforms \cite{elec-compare, goodenblockcomparison}. Both sets of papers show that Ethereum is a very popular, and, conservatively speaking, rather successful for commercial DLT software in this context.

In contrast with these observations, we found literature that highlights some of the key drawbacks of using Ethereum in a smart energy grid context \cite{smartgridblock, goodenblockcomparison}. These drawbacks include, but are not limited to:

\begin{itemize}
    \item Ethereum being a public blockchain means that additional efforts are required to anonymize customer data, which often means either using additional databases or having performance suffer from computational overheads caused by encryption.
    \item Ethereum is limited in how many transactions it can process per second. For small-scale systems this is not always problematic, but for a provincial or national-level power grid it could be a major hurdle to overcome.
    \item Ether (Ethereum's currency) fluctuates in value, which could make a very efficient system economically unsustainable almost overnight. 
\end{itemize}

Our results do showcase some of these drawbacks, especially when it comes to gas costs, which leaves us wondering why Ethereum is such a widespread production-level platform for energy solutions instead of a prototype-focused one. We suspect that energy costs in countries where successful instances were deployed may vary considerably from local energy costs, which has a noticeable impact on a system's viability. 

On a different note, we also found that, generally speaking, current blockchain-based energy tokenization systems opt to use fungible tokens \cite{powerledger, solarbankers, nrgcoin} instead of non-fungible ones. From a certification point of view, this means that fungible-token systems can only meaningfully discriminate between different categories of energy (e.g., renewable vs. non-renewable). However, the functional scope of such systems is hampered by that design decision, as is discussed in the Section \ref{sec:back}.

\section{Background}
\label{sec:back}
In this section, we will cover the minimum technical information required to understand how \zipzap~functions. 

\subsection{Tokenization Standards}
Tokenizing goods and services can be achieved with a wide variety of techniques. When it comes to \zipzap, however, it is most important to grasp the difference between fungible and non-fungible tokens (and their corresponding tokenization standards). At a glance, fungible tokens are interchangeable, whereas non-fungible tokens are unique---generally because the latter have rich metadata appended on an individual basis so they can be distinguished from one another. 

In the Ethereum network, the primary tokenization standard for fungible tokens is ERC-20 \cite{20-prop, erc-20-stats}. With it, creating alternative currencies is a straightforward process. Therefore if we wanted to exchange electricity under the assumption that customers only care about the amount of energy they get and its relative cost, then ERC-20 is more than adequate. In order to handle different energy sources, one could create multiple ERC-20 tokens. For example: there could be a token for photovoltaic energy, one for fossil-fuel energy and so on. 

Non-fungible tokens can be implemented using ERC-721. Adopting ERC-721 means that no two tokens are alike because they can be distinguished from one another based on the time and place where they were created, among other factors \cite{721-prop}. This is advantageous for the purposes of automating accounting tasks or allowing grid members to exclusively consume locally produced energy. The added metadata generates countless possibilities for giving users finer control over their energy production and consumption habits. Of course, storing and handling the metadata can quickly ramp up gas costs, which calls for careful consideration. 

With these trade-offs in mind, \zipzap~uses ERC-1155. ERC-1155 is a more recent tokenization standard that is backwards compatible with both ERC-20 and ERC-721 \cite{1155-prop}, so that both types of tokens can be used depending on the relevant use case---hence why we dubbed \zipzap~a semi-fungible system. Versatility is not the only appeal of this standard: it also allows tokens to be transacted in bulk, which helps save gas costs tremendously when compared to ERC-721 \cite{loglog-paper}.

Any smart grid application that wishes to support energy certification (especially if it requires energy traceability) would almost exclusively consider using a non-fungible tokenization standard (such as ERC-721 or ERC-1155) because it would otherwise be extremely difficult to discriminate between energy units of different provenance. If, however, the smart grid application only needs to allow users to transact energy units among each other with no regards for certification, then a fungible token standards (such as ERC-20 and ERC-1155) can be more than adequate. Either way, ERC-1155 gives us the flexibility to implement either without major code changes. Do note, however, that in all our prototypes we implemented non-fungible tokens for the sake of supporting certification in the future. That is also why we place an emphasis on appending metadata to each token: without it, it would not be possible to have meaningful records that could be audited for the sake of certification. This metadata is also useful for power grid maintenance and regulation as it could be used to spot areas and time periods that place higher/lower stress on the grid, as well as consumer/producer habits.

\subsection{Gas Costs}
Gas costs are a concept associated with Ethereum-based blockchain solutions. The complexity of computer programs is often measured in spatial complexity (the amount of data storage needed) and computational complexity (the amount of processing power needed). Since smart contracts are programs running on a distributed computer, gas costs are an amalgamation of the computational and spatial costs of hosting and running the program on the Ethereum network \cite{eth-yel}. Although some gas costs are only paid once (cost of deployment), ongoing costs (cost of transactions/operations) play the greatest factor in determining the viability of a distributed application.

Gas can be thought of as a unit of measurement for computational cost, whereas Ether (ETH) is the currency used to pay for any given amount of gas. By spending more ETH per unit of gas, one's transactions can be prioritized by the network, which results in a faster execution time (and vice-versa). Naturally, as both the average transaction fees and the price of Ether fluctuate, so too does the viablity of any one particular smart contract. Hence, Ethereum developers are almost always primarily concerned with minimizing the gas costs associated with their applications.

\subsection{Transactive Energy}

The concept of transactive energy refers to a collection of techniques used to manage energy exchanges. In contrast with the more centralized, hierarchical model used by its predecessors, transactive energy grids make full use of highly distributed networks where each of its members can directly interact with any other member. Using such a design provides granular control over each individual node and increases the volume of information that can be exploited to regulate the grid exponentially \cite{katipamula_transactive_2006}. In order to ensure informational accuracy and integrity, we believe that enriching transactive energy systems with blockchain technologies may be a viable approach for governing self-sustainable local markets. Solutions like \zipzap~could make energy grids more transparent and reliable, as well as enabling greater levels of automation through the use of smart contracts.

\section{Proposed Tokenization Solutions}
\label{sec:des}
Four different prototypes were developed for \zipzap, all with very different performance metrics. These differences can be entirely attributed to how much data each one of them stores and/or processes on-chain. The prototypes were named after weight categories, directly in proportion to their gas costs, with Heavyweight being the most expensive one and Weightless being the least expensive one. They are all homologous in terms of functionality, but vary in terms of extendability given how vastly they differ in terms of data handling. We provide Figure \ref{fig:comm-diag} for a quick overview of these differences.

\subsection{Zap}
\label{sec:Zap}
A \zap~is an ERC-1155 compatible non-fungible token that represents a given amount of energy. Each \zap~has the following metadata associated with it:

\begin{itemize}
    \item A history of its geographical location;
    \item A history of its owner’s accounts;
    \item The timestamp of its creation;
    \item The amount of energy it represents (in kWh);
    \item Power (in kW);
    \item Its estimated monetary value (in USD);
    \item Its generator’s ID number;
    \item The type of energy source exploited.
\end{itemize}

The most troublesome metadata fields are the two first ones listed, because they require, at a minimum, fixed-length arrays. Dynamic arrays and, to a much lesser extent, fixed-length arrays are quite costly in terms of gas. Heavyweight was able to save costs in this regard by limiting array size to 5, since realistically it is extremely unlikely a \zap~will ever be transferred more than once or twice since the energy losses incurred during transportation and storage are so large that electricity is most often consumed right away.

\zaps~are minted every five minutes and reflect the amount of energy created by a generator during that time window. When a user consumes electricity, they consume \zaps~until they have none left.

\subsection{Generalized Walkthrough}
\begin{figure*}[] 
\fbox{\parbox{\textwidth}{
	\centering
	\includegraphics[width=0.9\textwidth]{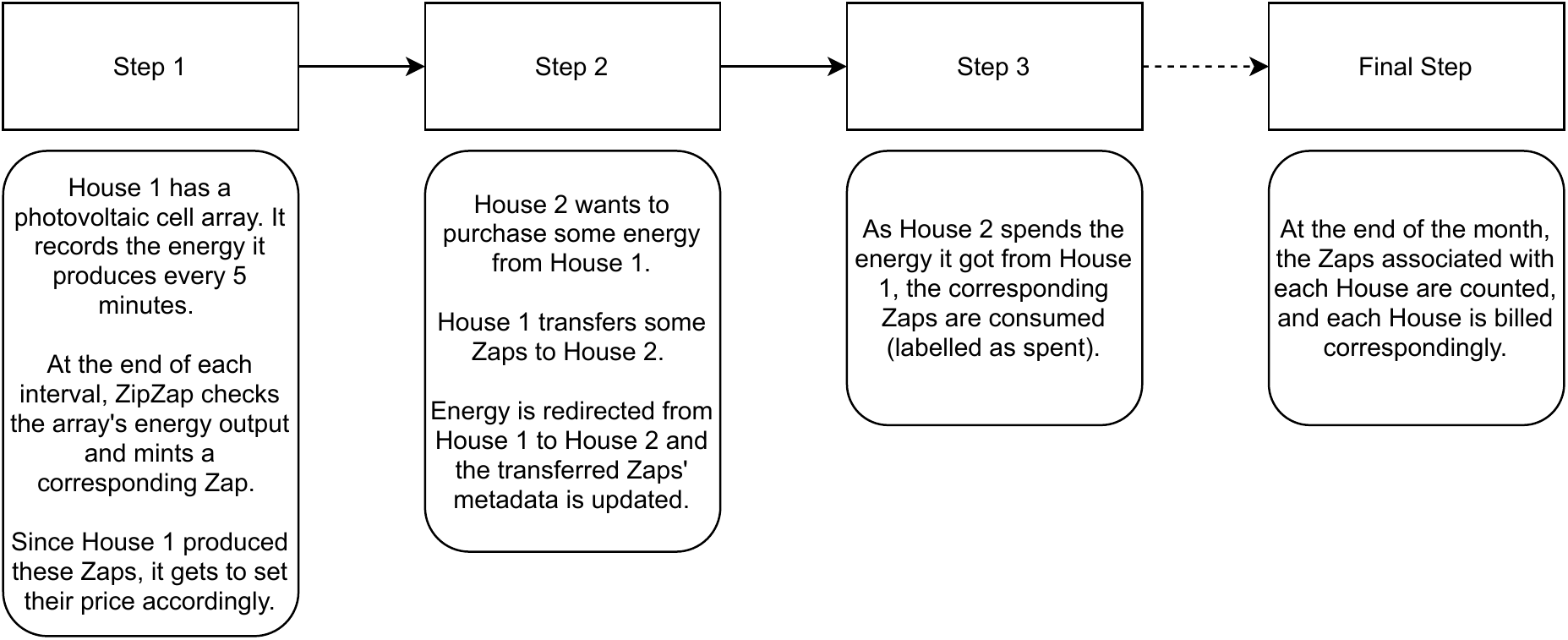}  
	}}
	\caption{ZipZap Walkthrough}
	\label{fig:zapflow}
\end{figure*}

Regardless of the prototype considered, \zipzap~works roughly as shown in Figure \ref{fig:zapflow}. \zaps~are used to track all energy exchanges throughout the system. Whenever an operation is conducted on a \zap, its metadata has to be updated correspondingly. Note that although steps 1-3 take place multiple times per day, customers are only billed at the end of each payment cycle, at which point they directly pay the producer of each \zap~they held during that period. If customers purchased enough \zaps~to cover their consumption, then the cycle repeats without any further steps. If, however, customers did not purchase enough \zaps~to cover their energy consumption, then they are charged for the difference by the de-facto large-scale provider of the energy grid, which would then transfer an equivalent number of \zaps~to the customer in question. 

It is also worth noting that in its current stage, \zipzap~does not allow price-bidding, instead relying on the price originally set by a \zap's creator upon minting. However, such a system is envisioned as part of future extensions to this research.

\subsection{Logistical Considerations}
\label{sec:logcon}

As far as our prototypes are concerned and using Figure \ref{fig:houses} as a reference, network nodes could be hosted by an energy company's local transformers and/or by local libraries, banks or any company willing/responsible enough to look after them. Energy companies themselves would be more than happy to assume that responsibility given that, in the case where a local network does not produce enough energy to cover its own demand, the energy company in charge of the node would settle the difference using its own energy supply and corresponding rates. Note, however, that balancing generation and consumption is a complex problem of its own and not the main focus of our research. All the metadata associated with each token may help form smarter solutions to this end, but we do not offer an implementation of such a regulatory system.

\subsection{\heavy}
\label{sec:overheavy}
Heavyweight was the first prototype developed. Practically speaking, all of the metadata for each and every \zap~is stored entirely on-chain. Consequentially, gas costs are quite steep. Attempts were made early in development to integrate not only \zaps, but also batteries and generators into the system, but gas costs quickly resulted in scoping out the two latter, since the additional spatial requirements and their impact on gas costs were far too great. 

It is also worth noting that this is the only prototype that defines a \zap's metadata using standard Solidity data types such as fixed-sized integer arrays. All the other prototypes (including \noweight) store and/or handle each \zap's metadata as a JSON file that has to be carefully parsed whenever it is read or modified. This is a gas-cost saving measure since computational complexity did not have as great an impact as spatial complexity in our tests. 

\subsection{\feather}
\label{sec:overfeather}
Featherweight is meant to represent the absolute lowest gas costs one could reasonably expect to get out of an Ethereum-based system with the same functionality as Heavyweight. A \zap's metadata is stored off-chain, with only the hash of the aforementioned metadata being stored on-chain. This changes how transfers are handled. Since the smart contract cannot directly update the metadata after a transfer takes place, the new owners of the affected \zaps~have to manually call a ``modify \zap" function to update the metadata hash of the affected \zaps~after a trade takes place. Therefore, the total cost of a transfer is the sum of the gas costs from the transfer itself and from the ``modify \zap" function. 

\subsection{\light}
\label{sec:overlight}
Featherweight also stores only \zaps' metadata. However, in contract with Lightweight, its functions require that the full \zap~metadata be sent to the smart contract for all \zap~related operations. The first step for any such operation is verifying the validity of the sent metadata using the stored hash. Once validated, if a change must take place in the metadata, the contract parses it and modifies it as needed before updating the stored hash and sending back the modified metadata. Therefore, although it stores the same amount of information as Featherweight, Lightweight has a higher degree of automation and is more easily extended to include more complex operations.  

\subsection{\noweight}
\label{sec:noweight}
\noweight~is based on a different blockchain platform: Quorum. Since Quorum is a fork of Ethereum, it is capable of using the same programming language and libraries as our other prototypes. Thus, it is the closest way to compare the performance of our Ethereum prototypes to a gas-free system. In lieu of gas, the costs associated with running a gas-free system are predominantly dependant on the number of full nodes used by the network, which in turn determines hardware purchasing costs, operational costs and maintenance costs. These cannot be meaningfully estimated without intending to conduct a full deployment because all of the aforementioned costs fluctuate violently depending on legislation, local electricity costs, hardware availability and many other factors like required fault tolerance and system size. 

Other than having no gas costs, \noweight~performs almost identically to \light~because the former was built using the latter as a basis and, consequentially, operates virtually in the same way--all functions have the same complexity, with differences in time performance ultimately depending on node propagation times. It is worth noting that, in the interest of fairness, since all Ethereum tests were conducted on a local test chain (where node propagation is not a factor), all Quorum tests were conducted on a single local test node.


\section{Results and Discussions} 
\label{sec:res}
We begin this section by defining our laboratory environment. We then provide, for each prototype on Ethereum: 
\begin{itemize}
    \item their gas costs.
    \item an analysis of the aforementioned costs.
    \item some basic time-performance metrics and how they relate to the previous two points.
\end{itemize}

Subsequently, we present gas cost and time performance comparisons between our Ethereum prototypes, followed by cost and performance analysis of our Quorum prototype. Finally, we close the section by discussing the cost viability of all solutions. 

\subsection{Setup and Parameters}

We decided to use Ethereum as a development platform not only because of its maturity and flexibility, but also because it has readily available tokenization libraries (notably from OpenZeppelin) that considerably sped up prototyping. Thus, the first three prototypes were developed using Solidity. After preliminary gas cost analysis, it was evident that exploring a gas-less alternative was imperative. In order to still have access to the same tokenization libraries and have as close a comparison as possible to our original set of prototypes, we used Quorum \cite{quorum}, an Ethereum fork, to implement \noweight.

All prototypes were compiled and tested using Truffle Suite on a local test network. Hardware-wise, all tests were ran from an x86\_64 Linux machine using version 5.9.16-1 of Manjaro. The computer in question was built with an 8-core AMD Ryzen 7 3700X CPU and 16GB of RAM.

\subsection{Gas Results and Comparison}
\begin{figure}[] 
\fbox{\parbox{\columnwidth}{
\centering
	\begin{subfigure}{0.8\columnwidth}
		\centering
		\includegraphics[width=\columnwidth]{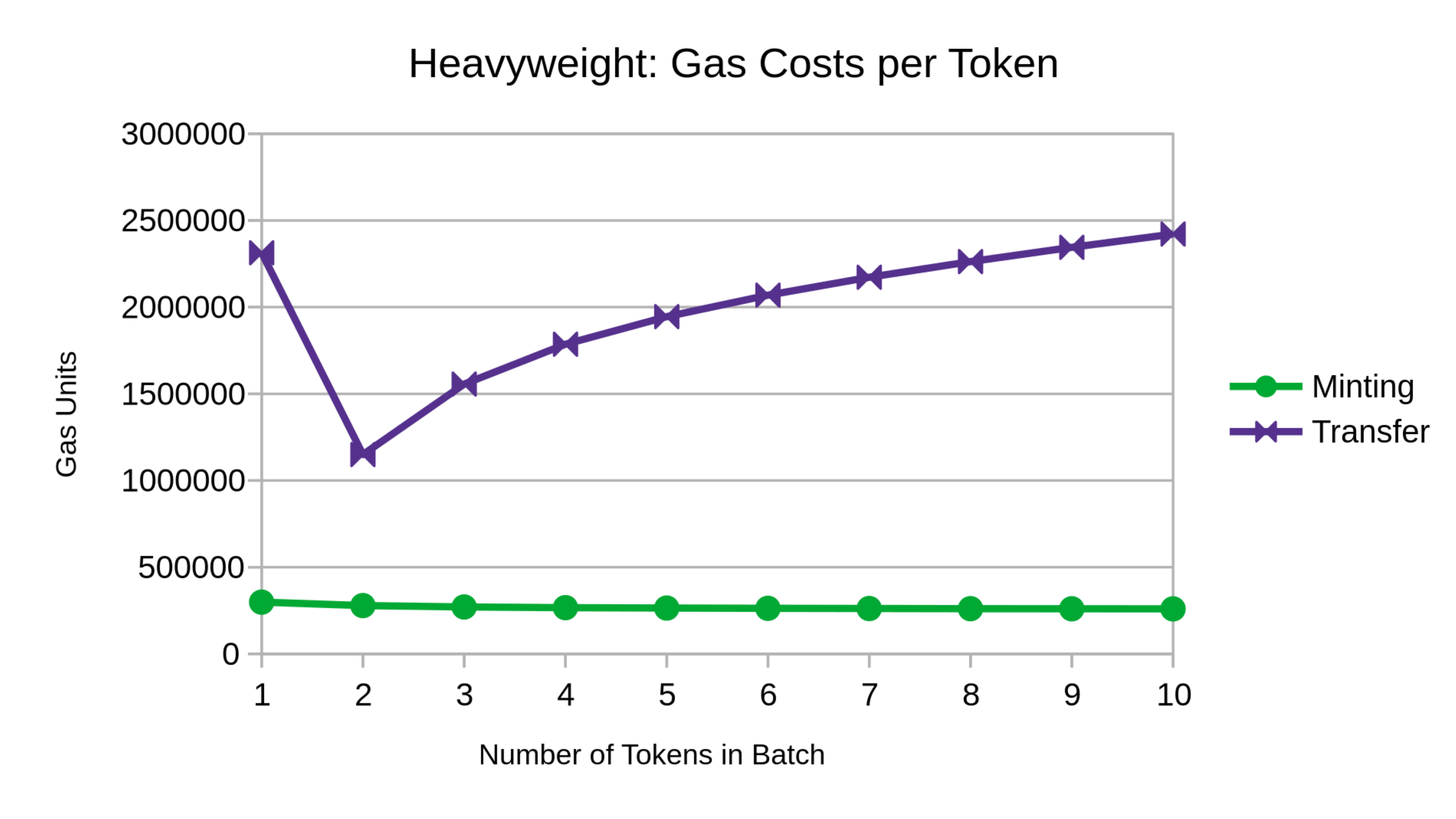}  
		\caption{}
		\label{fig:hgct}
	\end{subfigure}
	
	\begin{subfigure}{0.8\columnwidth}
		\centering
		\includegraphics[width=\columnwidth]{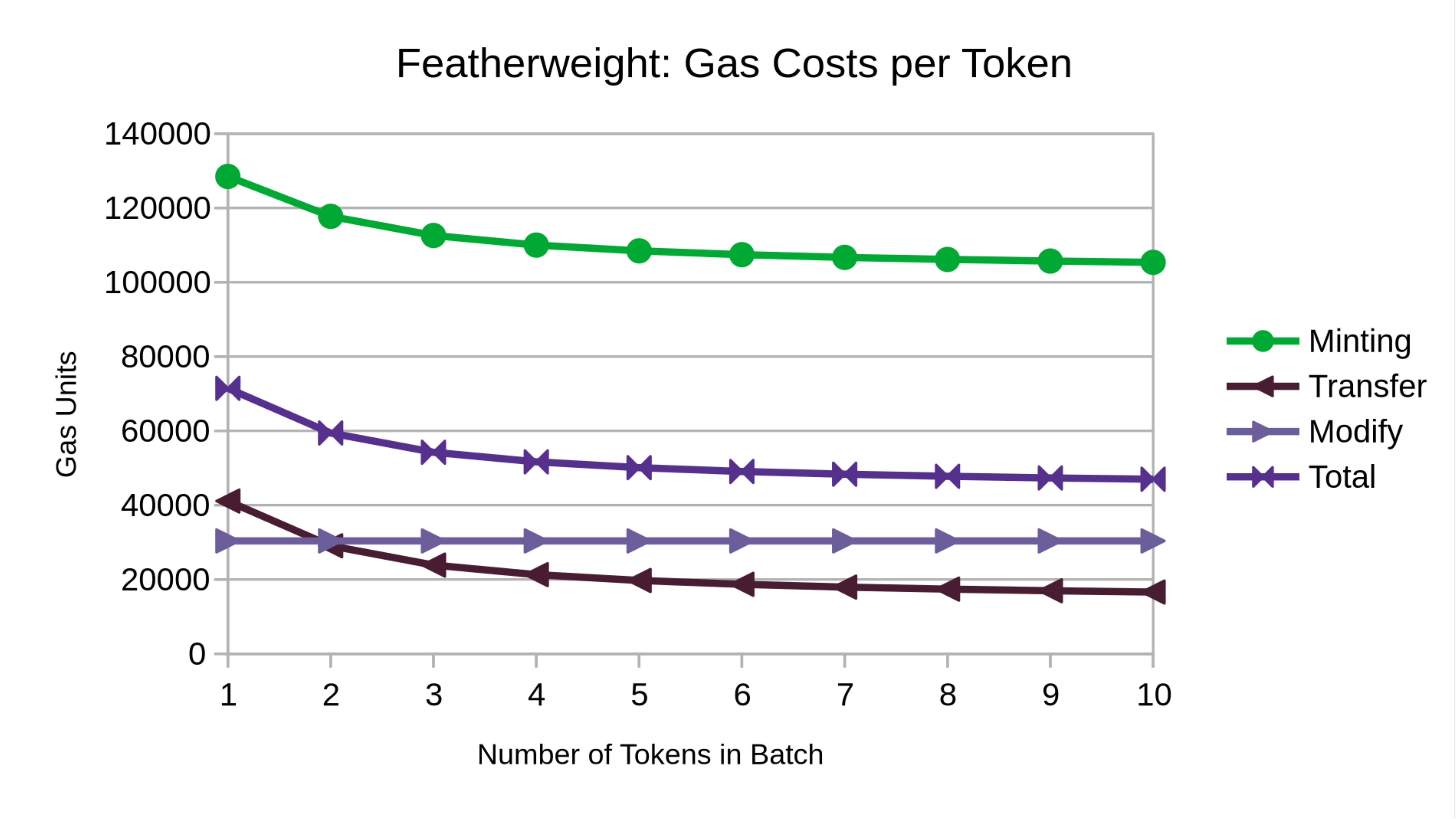}  
		\caption{}
		\label{fig:fgct}
	\end{subfigure}
	
	\begin{subfigure}{0.8\columnwidth}
		\centering
		\includegraphics[width=\columnwidth]{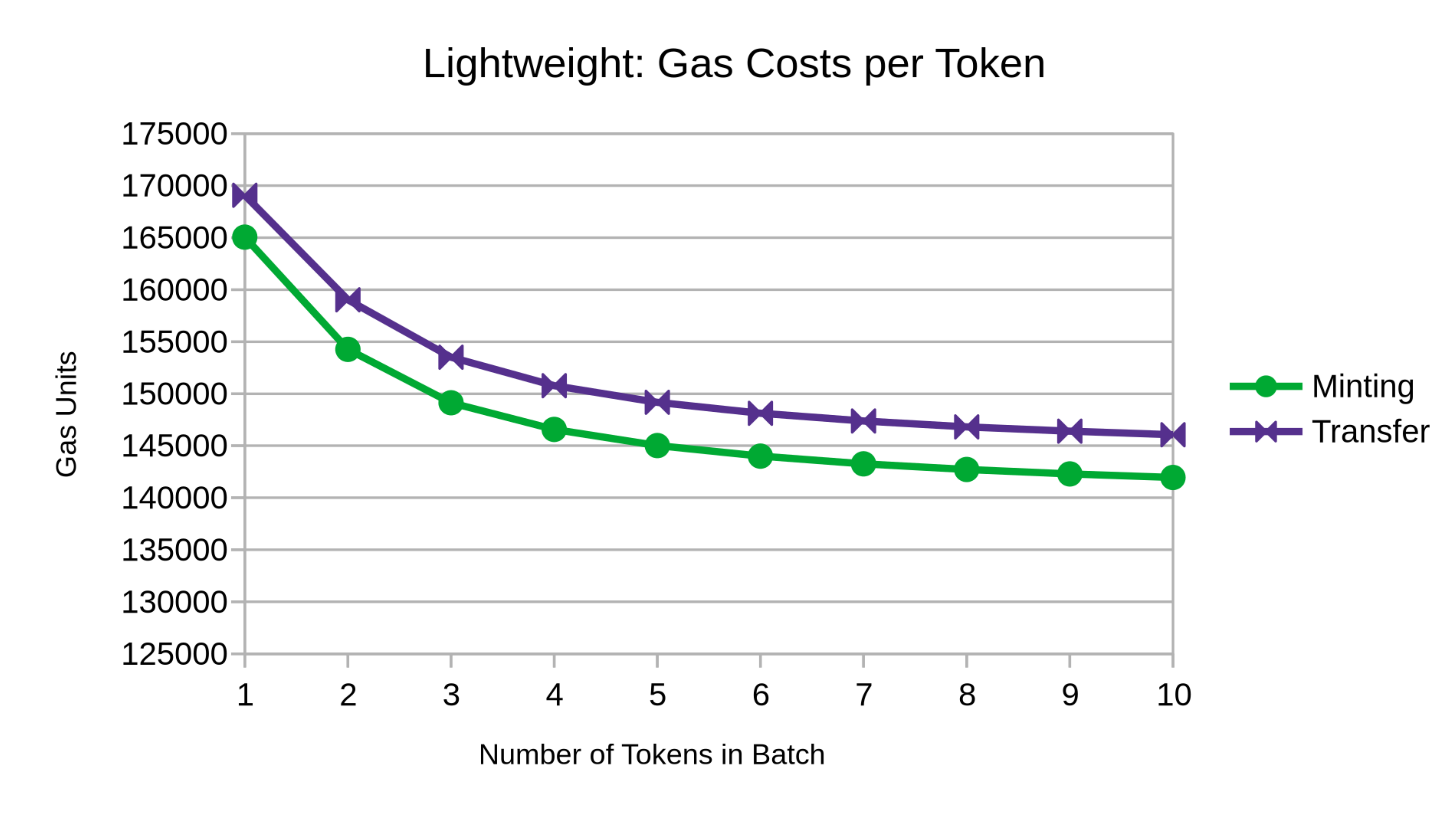}  
		\caption{}
		\label{fig:lgct}
	\end{subfigure}
	}}
	\caption{Gas Costs per Prototype}
	\label{fig:gc}
\end{figure}

%

\heavy's gas performance (as shown in Figure \ref{fig:hgct}) deviated from the norm in terms of bulk transformation savings. Whereas one would expect a constantly decreasing asymptotic curve (like the ones in Figure \ref{fig:lgct}), our results show a sharp decrease in gas costs, followed by a constantly increasing asymptotic curve. This is due to implicit on-chain operations out-scaling the gas savings associated with performing transactions in bulk. Namely, when any number of \zaps~are transferred, the system has to update the metadata of each of them to reflect the change in ownership and address. This specific operation carries a computational cost of $O(N)$, whereas bulk savings are approximately in the order of $1/O(log(N))$, as shown by the gas costs of \feather~and \light.

%

\feather~is an interesting edge case. As noted in Section \ref{sec:overfeather}, the total cost of a transfer in this context is the sum of the gas costs from the transfer itself and from the ``modify \zap" function, hence the additional lines in Figure \ref{fig:fgct}. Beyond being the lightest Ethereum-based prototype, the main peculiarity of \feather~is that it is the only prototype where minting is more expensive than transferring. This is because, despite having to add the costs of modification, all transfer-related sub-operations (like updating a \zap's information) have to be manually triggered, so there are far fewer checks, validations and automatic triggering of functions for every transfer. This results in considerable gas savings for transferring tokens, since the operation is simplified. 

%

\subsection{\light~Gas Results}
\light~is a happy medium between \feather~and \heavy, handling metadata in a similar manner to most blockchain applications. It is the most viable of all three Ethereum prototypes, not strictly because of its gas metrics, which are slightly inferior to those of \feather,  but because its processing of metadata allows it to be easily extended in terms of functionality. In other words, when compared to \feather, it would be almost trivial to extend \light~to include more features because it actually reads and modifies the metadata related to each token even if it only stores the hash. \heavy, on the other hand, has all the information it would need for many future use cases one could consider, but is limited by the already disproportionate costs of storing all that data.

\subsection{Gas cost comparison on Ethereum}

\begin{figure}[] 
    \fbox{\parbox{\columnwidth}{
	\begin{subfigure}{\columnwidth}
	    \centering
	    \includegraphics[width=\columnwidth]{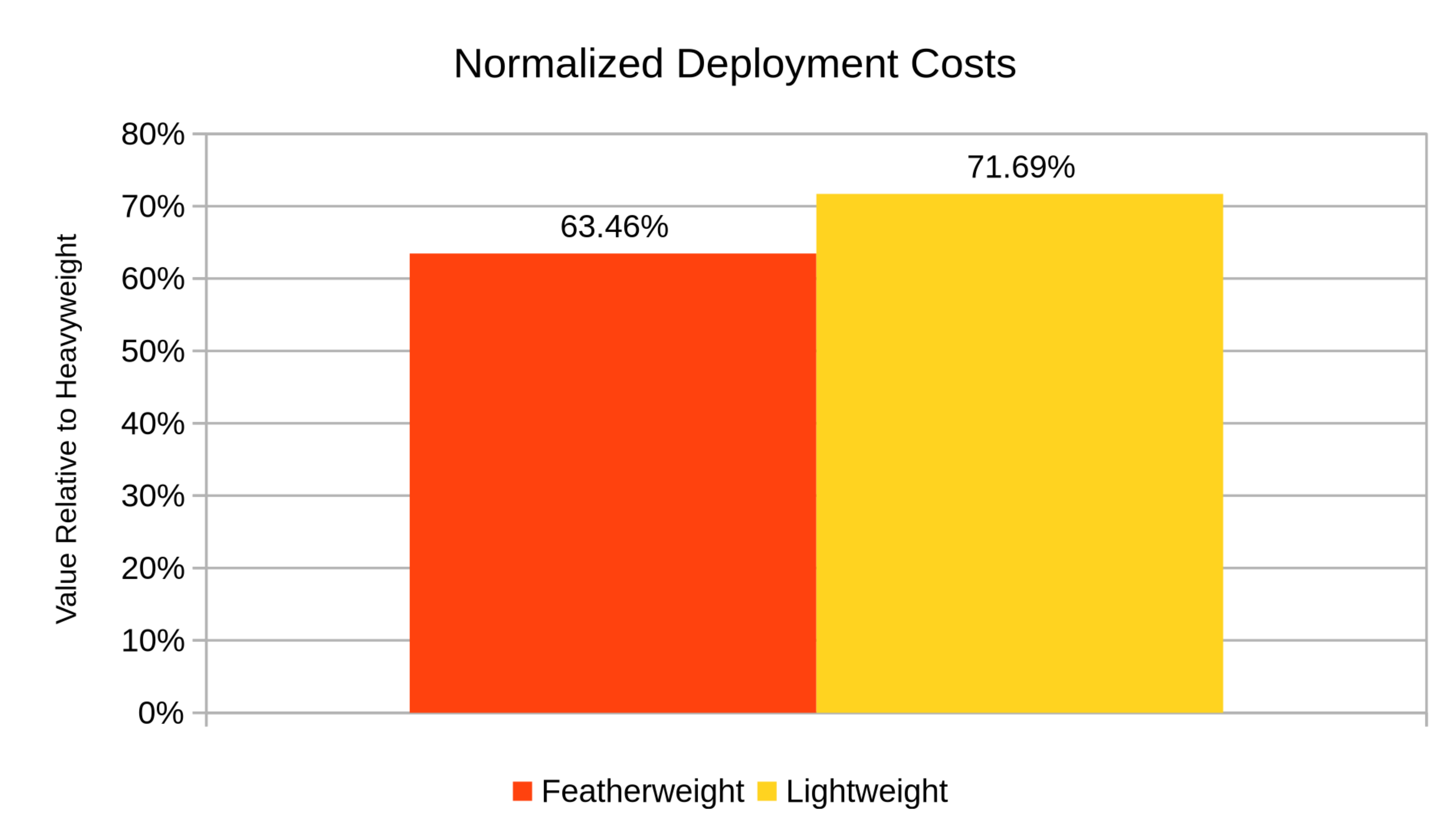}  
	    \caption{}
	    \label{fig:ndc}
	\end{subfigure}
	
	\begin{subfigure}{\columnwidth}
	    \includegraphics[width=\columnwidth]{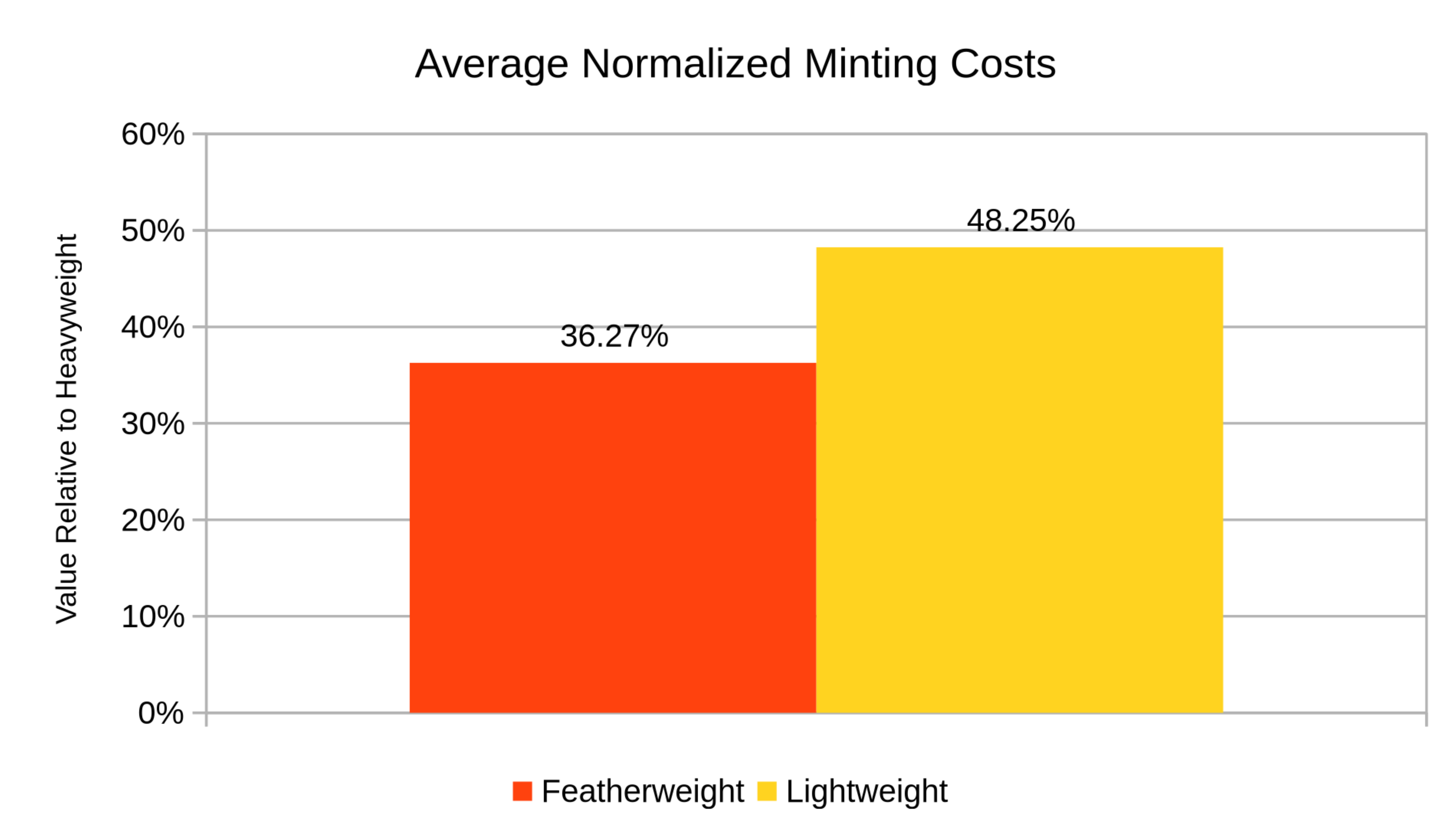}  
	    \caption{}
	    \label{fig:anmc}
	\end{subfigure}
	
	\begin{subfigure}{\columnwidth}
	    \includegraphics[width=\columnwidth]{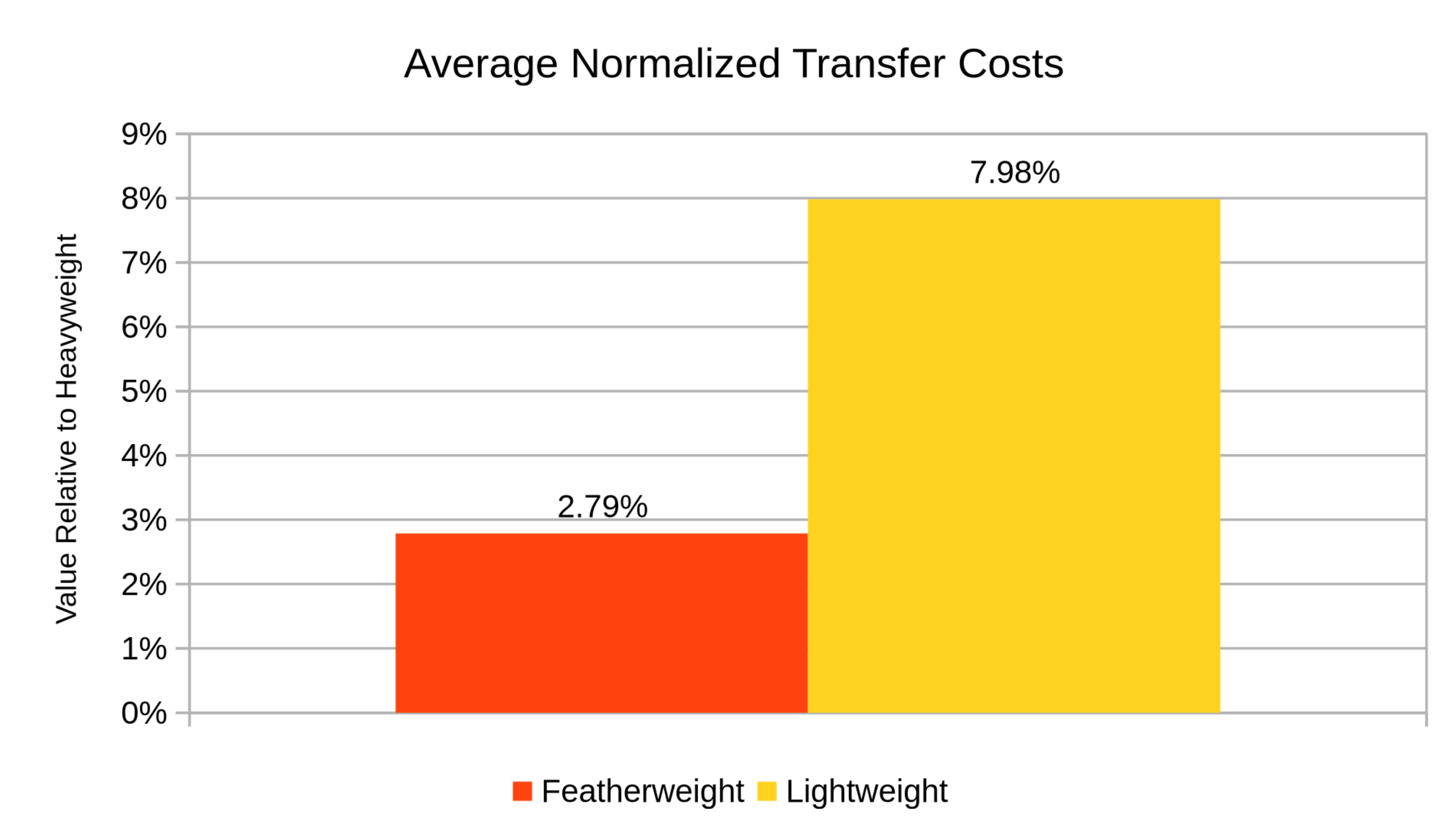}  
	    \caption{}
	    \label{fig:antc}
	\end{subfigure}
	}}
	\caption{Gas Cost Comparison}
	\label{fig:gas2}
\end{figure}

Figure \ref{fig:gas2} shows how substantial the gas cost reductions are. All of the graphs shown therein are normalized in relation to \heavy's results. From these graphs, we can infer that the gas reductions attributed to storing metadata off-chain vary between approximately 40 to 30 percent when it comes to deployment. However, the biggest, most impressive savings come from recurring operations, where the cost associated with minting tokens is reduced between 50 and 60 percent and, incredibly, between 97 to 90 percent when it comes to transferring tokens.

\subsection{Time Performance Comparison on Ethereum}
\zipzap's latency requirements and daily rate of operations are tied to one peculiarity of the power counting equipment currently installed throughout Quebec: power readings are recorded on five minute intervals.

\begin{figure*}[] 
\fbox{\parbox{\textwidth}{
\centering
	\begin{subfigure}{0.8\columnwidth}
	    \includegraphics[width=\columnwidth]{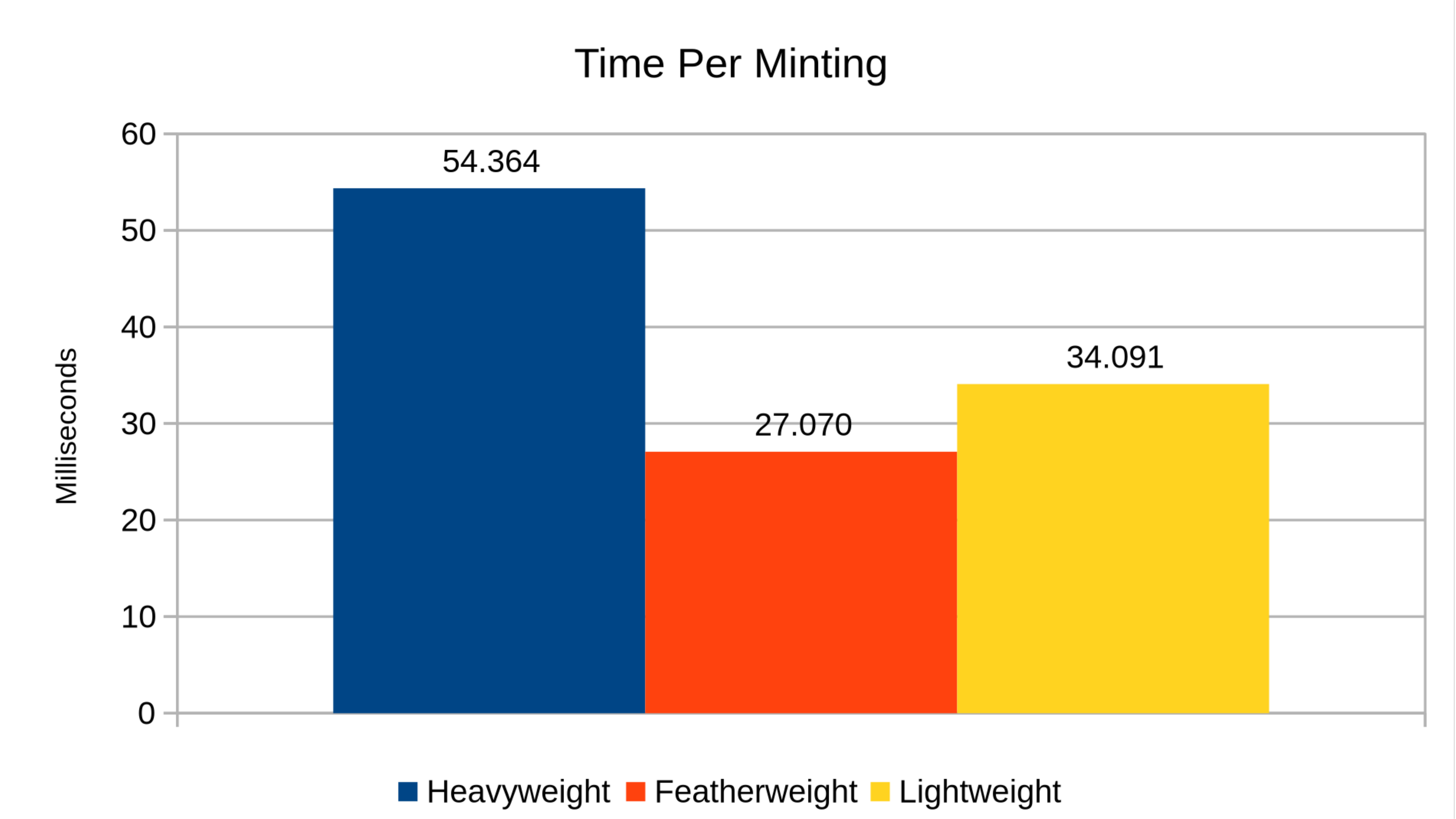}  
	    \caption{}
	    \label{fig:tpm}
	\end{subfigure}
	\begin{subfigure}{0.8\columnwidth}
	    \includegraphics[width=\columnwidth]{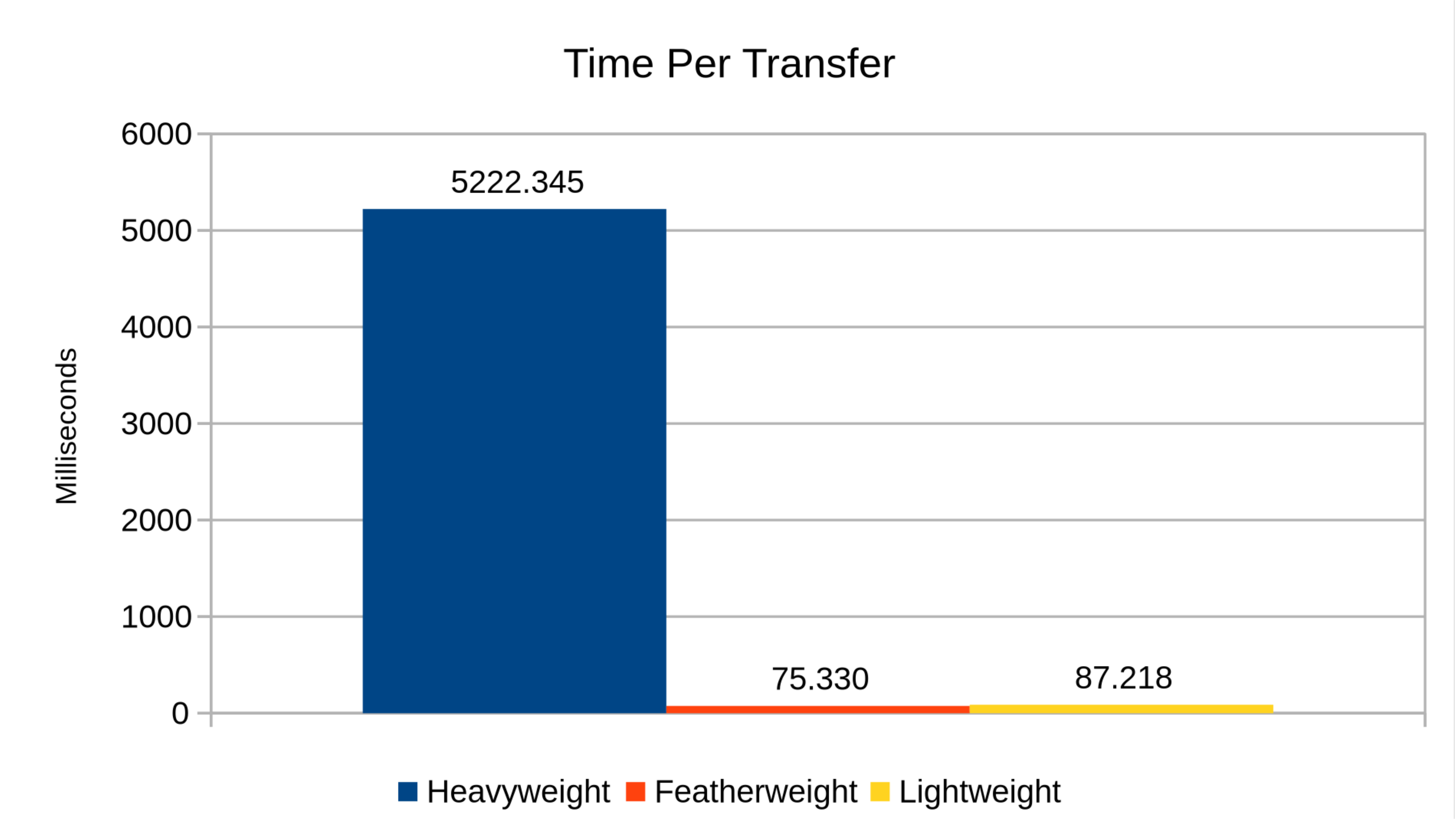}  
	    \caption{}
	    \label{fig:tpt}
	\end{subfigure}
	
	\begin{subfigure}{0.8\columnwidth}
	    \includegraphics[width=\columnwidth]{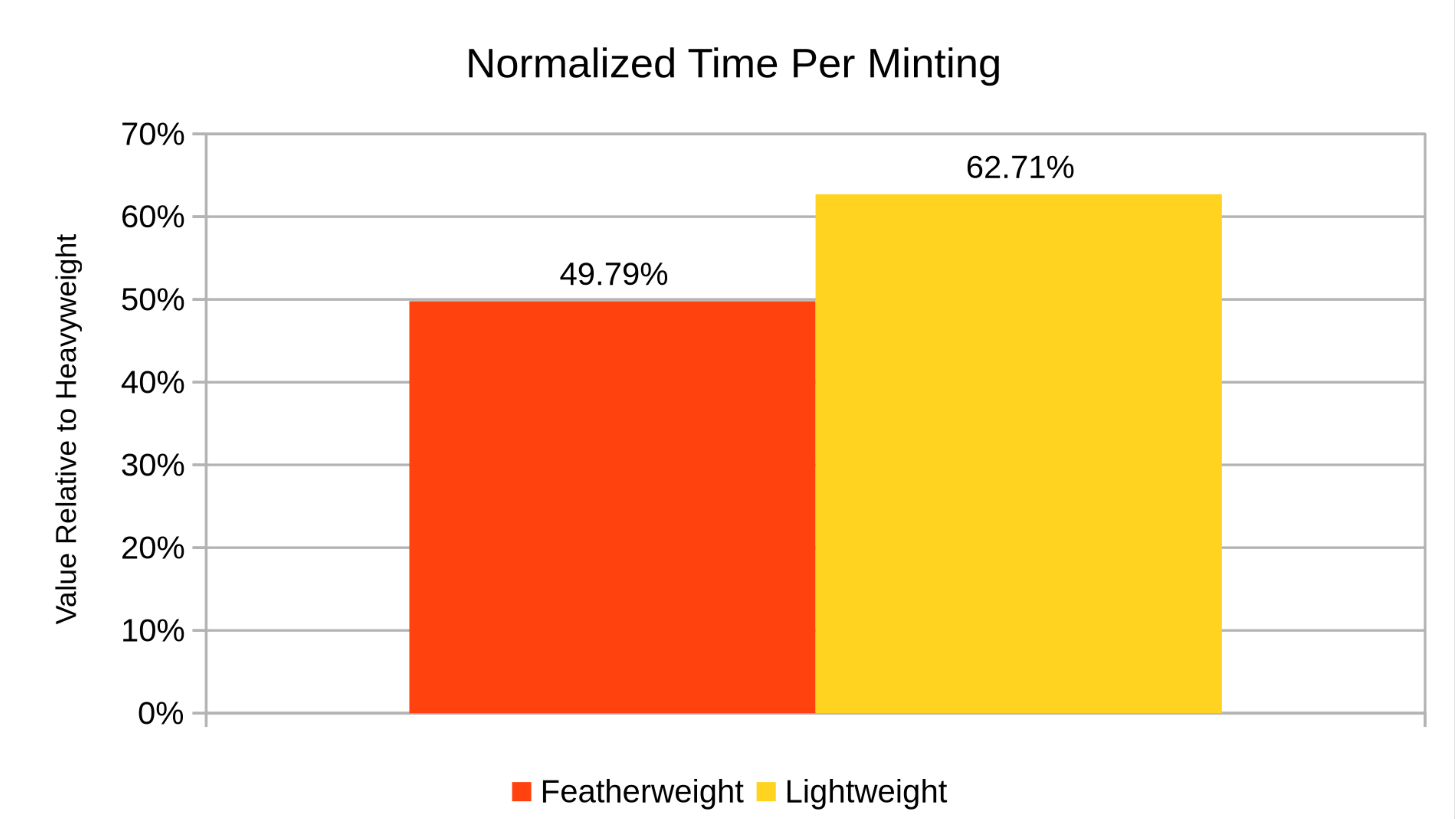}  
	    \caption{}
	    \label{fig:ntpm}
	\end{subfigure}
	\begin{subfigure}{0.8\columnwidth}
	    \includegraphics[width=\columnwidth]{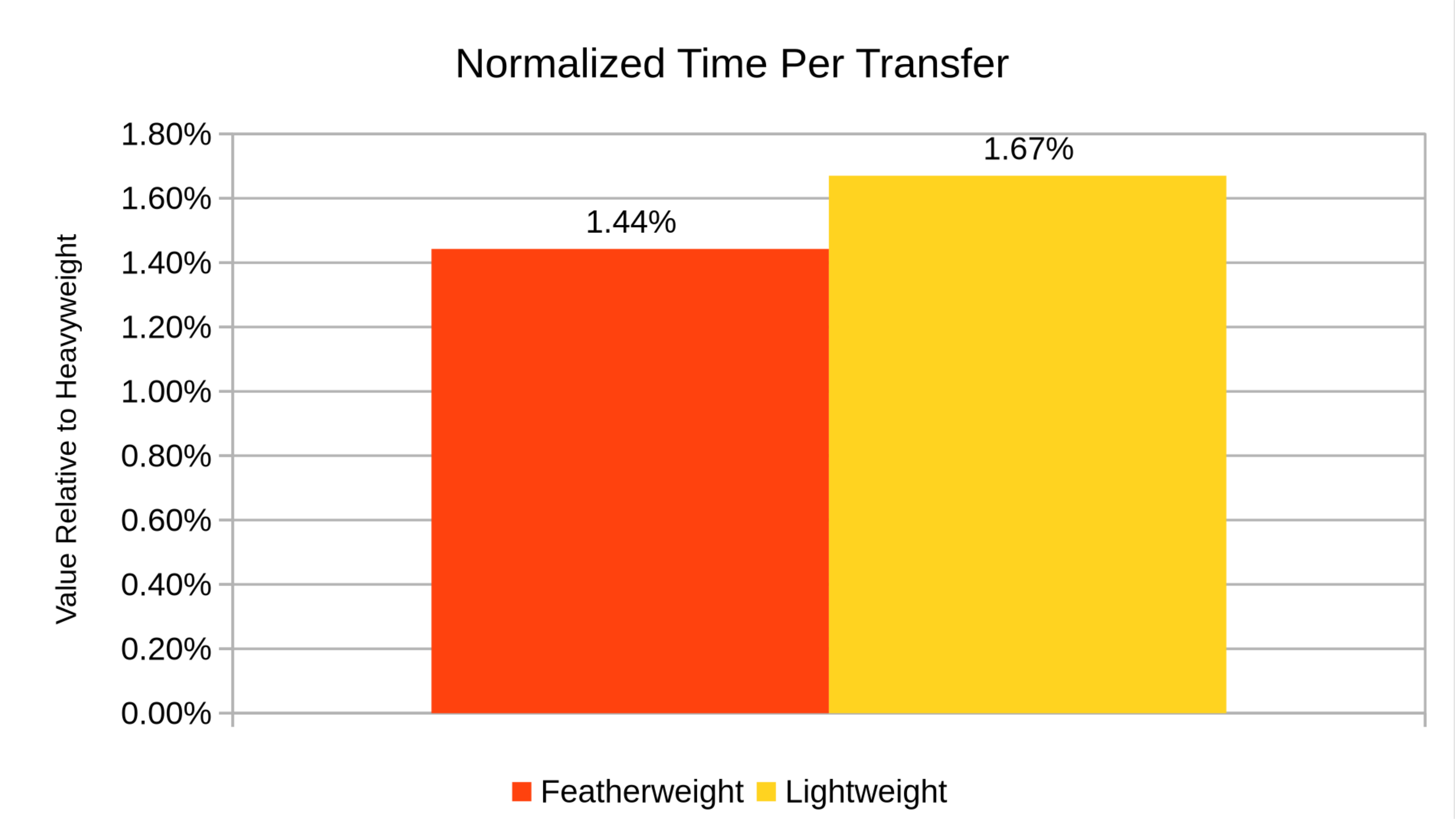}  
	    \caption{}
	    \label{fig:ntpt}
	\end{subfigure}
	}}
	\caption{Time Performance Comparison}
	\label{fig:time}
\end{figure*}

This means that all households must mint and/or transfer \zaps~within that same five minute window. Assuming every household in the system generates some form of energy, this translates to, at a minimum, 288 mintings per household, per day. This would not be a problem if they could be handled in bulk, but because they are spaced out in time and on a per-household basis, it is difficult to make the most out of the gas savings provided by ERC-1155 tokenization. Regardless, Figure \ref{fig:time} shows that all three prototypes manage to meet the five minute requirement (with massive latency reductions recorded for both \feather~and \light) as far as processing time alone is concerned. It does not take into account block confirmation time, which would add an additional 3-5 minute window. In most cases, this would still result in an acceptable performance, but changes in network demand are difficult to anticipate and would impact the confirmation delay. This is actually another advantage that a Quorum-based implementation would have over an Ethereum-based one: block confirmation time would be infinitesimal in comparison (milliseconds instead of minutes). 

\subsection{\noweight~Cost and Performance Analysis}
In terms of gas units, \noweight~has the same performance as \light~(see Figure \ref{fig:lgct}). Incidentally, \noweight~also has the same time performance as \light~(see Figure \ref{fig:time}). This is because the underlying code-base for both of these prototypes is virtually identical, so all operations are essentially the same. This was done in an effort to provide as close a comparison as possible between the two blockchain platforms considered. 
\begin{table*}[] 
\caption{Cost Analysis of \zipzap's \feather~and \light~Versions}
	\renewcommand{\arraystretch}{1.3}%
	\scriptsize
	\centering
	\resizebox{\textwidth}{!}{
			\begin{tabularx}{\textwidth}{| >{\centering\arraybackslash}X | >{\centering\arraybackslash}X || >{\centering\arraybackslash}X | >{\centering\arraybackslash}X | >{\centering\arraybackslash}X | >{\centering\arraybackslash}X | >{\centering\arraybackslash}X | }
				\hline
				 \textbf{Prototype}& \textbf{Operations} & \textbf{Deployment} & \textbf{Minting} &	\textbf{Safe Transfer} & \textbf{Batch Minting (10 Tokens)} & \textbf{Batch Transfer (10 Tokens)}
				\\ 
				\hline 
				\multirow{5}{*}{\textbf{\feather}} & \textbf{Gas Units Used} & 3702977 & 158483 & 71456 & 1053673 &	469711 
				\\ 
				\cline{2-7}
				& \textbf{Fast Cost (Gwei)} & 111089310 & 4754490 & 2143680 & 31610190 & 14091330
				\\ 
				\cline{2-7}
				& \textbf{Fast Cost (USD)} & 26.66	& 1.14 & 0.51 & 7.59 & 3.38 
				\\ 
				\cline{2-7}
				& \textbf{Standard Cost (Gwei)} & 96277402 & 4120558 & 1857856 & 12212486 & 12212486 
				\\ 
				\cline{2-7}
				& \textbf{Standard Cost (USD)} & 23.11 & 0.99 & 0.45 & 2.93 & 2.93 
				\\ 
				\hline 	
				\multirow{5}{*}{\textbf{\light}} & \textbf{Gas Units Used} & 4592780 & 165052 & 169077 & 1419443 & 1460545
				\\ 
				\cline{2-7}
				& \textbf{Fast Cost (Gwei)} & 137783400 & 4951560 & 5072310 & 42583290 & 43816350
				\\ 
				\cline{2-7}
				& \textbf{Fast Cost (USD)} & 33.07 & 1.19 & 1.22 & 10.22 & 10.52 
				\\ 
				\cline{2-7}
				& \textbf{Standard Cost (Gwei)} & 119412280 & 4291352 & 4396002 & 37974170 & 37974170
				\\ 
				\cline{2-7}
				& \textbf{Standard Cost (USD)} & 28.66 & 1.03 & 1.06 & 9.11 & 9.11 
				\\ 
				\hline 	
				\end{tabularx}}
	\label{tab:costan2}
\end{table*}
As mentioned previously, \noweight~does not incur gas costs (even though technically it does spend gas units), because we shoulder the costs of running the Quorum network ourselves. Instead, the monetary costs associated with running \noweight~depend on the number of nodes required by the system, which would be partially estimated based on:

\begin{itemize}
    \item The number of households/members that are part of the smart grid
    \item The level of fault tolerance required by law or otherwise desired by the market
    \item The costs of renting or owning adequate servers for each node
\end{itemize}

Obviously, the cost of such a system can fluctuate tremendously, so instead of a thorough estimation, we propose a simple estimate using the same scale as our tests: one node servicing one small neighbourhood. Our first assumption is that a Linux VPS with 16 GB of RAM, 320GB of storage and a 4-core processor can be had for 80 USD/month \cite{amazonserver}, regardless of how many operations the system performs. We will now contrast that number with \light's monthly costs. Let us assume that safe transfers occur with the same frequency as minting (once per household, every 5 minutes) and that all such operations are conducted individually instead of being batched, resulting in a worst-case scenario. Then, if we have just a measly 2 houses in the neighbourhood and omit deployment costs, we estimate \light's monthly costs using Table \ref{tab:costan2} as:

\begin{itemize}
    \item $5 \times 12 \times 24 = 288$ mintings and $288$ transfers per house, per day.
    \item $288 \times 1.03 = 296.64$ USD spent in minting fees, per house, per day
    \item $288 \times 1.06 = 305.28$ USD spent in transfer fees, per house, per day
    \item $296.64 \times 30 \times 2 + 305.28 \times 30 \times 2 = 36115.20$ USD spent monthly, system-wide. 
\end{itemize}

Clearly, there is no comparison whatsoever between Ethereum and Quorum in terms of gas costs. Having said that, we nonetheless would like to note that once Ethereum 2.0 is fully released, these metrics may change drastically \cite{eth2} and may be worth revisiting.

\subsection{Cost Viability Discussion for All Solutions}

In this section we make the same assumptions related to real-world gas costs \cite{eth-cost} that we made in Section \ref{sec:res}, but we also assumed that the energy cost for a single kWh hovered around 30 cents. Under these assumptions, the gas results obtained for both \light~and \feather~were processed to produce Table \ref{tab:costan2}. The table shows that despite our promising results, even the lightest of our Ethereum-based prototypes can only be economically viable if the amount of energy stored by each \zap~exceeds one kWh by several orders of magnitude. This is of course hardly reasonable since each generator in the system must emit a \zap~every five minutes as per our requirements, and we are dealing with small generators that supply a single-household instead of provincial hydroelectric power plants. One way to sidestep these costs while still using Ethereum as a development platform would be to create a future-market-based implementation so that \zaps~can be transferred in bulk once or twice per day. Regardless of any such optimizations, at this point it seems obvious that Quorum is much better suited for smart grids than Ethereum.

\subsection{Future Work}
\label{sec:lesful}
Whereas DLTs are almost exclusively used in conjunction with other types of databases, we decided to explore the performance of a fully on-chain system. Common sense dictates that on-chain systems are more expensive from a gas point of view, but that does not mean that one such system is altogether not viable. Previous research has shown at least one combination of circumstances where an on-chain system is comparable in term of ongoing monetary expenses to a currently used system and boasts significant functional and security advantages \cite{loglog-paper}. Therefore, we encourage others to explore different contexts where on-chain systems may also be viable.

Developing a system similar to \zipzap~using a different tokenization standard is another attractive lead. For example, an ERC-998 solution would have a significantly different architecture and gas performance when compared to ours due to the former's capability for composing tokens, since generators and energy tokens are logically connected in a composition-like manner. It may thus be possible to not only render the system more efficient, but to also increase its scope by considering other tokenization standards.

Similarly, it may be worthwhile to explore other blockchain platforms beyond Ethereum and its many forks to better measure the impact that different network characteristics have on homologous distributed applications. 

\section{Conclusion}
\label{sec:conc}
We began work on this paper in hopes of increasing the transparency, informational integrity and degree of automatization within smart energy grids by developing a blockchain-based solution that would seamlessly integrate with them. Developing a fully on-chain solution proved to be a difficult task given the current state of DLTs. 

Despite our most efficient Ethereum-based prototype reducing baseline gas costs by more than 97\%, our results demonstrate that in order for an Ethereum-based system to be a viable solution for energy tokenization, time-related requirements must be relaxed either by increasing the allowed latency per operation, or by reducing the number of daily operations. On a similar note, although Ethereum-based versions of \zipzap~are currently not economically viable at a small scale, large-scale energy suppliers may still find the solution very attractive due to the much larger energy quantities generated and transferred at any time interval. This is a very interesting result that stands in contrast with the overwhelming popularity of Ethereum as a development platform for similar applications \cite{smart-examples, smart-categories}.

On the other hand, gas-less blockchain systems such as the one used by \noweight~are more attractive to small scale use cases and can in fact be commercially viable depending on how fault-tolerant the network needs to be. Naturally, maintaining more nodes is the price to pay for stability, but those costs could potentially be offset by incentivizing network members to maintain their own nodes, depending on the business model used.







%

\bibliography{bibliography}{}
\bibliographystyle{IEEEtran}


\end{document}